\documentclass[final,5p,times,twocolumn]{elsarticle}

\usepackage{etoolbox}
\usepackage[utf8]{inputenc}

\usepackage{mathrsfs,mathtools,bbold}
\usepackage{amsmath,amssymb}
	\allowdisplaybreaks[1]
\usepackage{amsfonts}
\usepackage{float}
\usepackage[colorlinks,linktocpage]{hyperref}
\usepackage[dvipsnames]{xcolor}
\usepackage{physics}
\usepackage{slashed}
\usepackage{bm, cancel}
\usepackage[flushleft]{threeparttable}
\usepackage{multirow}
\usepackage{booktabs}
\usepackage{verbatim}

\usepackage[normalem]{ulem}

\hypersetup{colorlinks=true,linkcolor=blue,citecolor=blue,urlcolor=blue}

\biboptions{numbers,sort&compress}
\journal{\color{gray}\phantom{arxiv-hep}}

\begin{document}

\begin{frontmatter}

\title{Predicting outcomes of electric dipole and magnetic moment experiments}

\author[label1]{V. G. Baryshevsky}
\ead{v\_baryshevsky@yahoo.com}

\author[label2]{P. I. Porshnev}
\ead{pporshnev@gmail.com}

\address[label1]{Institute for Nuclear Problems, Belarusian State University, Minsk, Belarus}
\address[label2]{Past affiliation: Physics Department, Belarusian State University, Minsk, Belarus}

\begin{abstract}
 The anomalous magnetic and electric dipole moments in spin motion equation acquire pseudoscalar corrections if the $T(CP)$-noninvariance is admitted. It allows to explain the discrepancy between experimental and theoretical values of muon $(g-2)$ factor under assumption that the pseudoscalar correction is the dominant source of this discrepancy.
\end{abstract}



\end{frontmatter}

\section{Introduction}
The long-standing discrepancy \cite{aoyama_anomalous_2020, abi_et_al_measurement_2021} between experimental and theoretical data for muon $(g-2)$ factor is a potential indication of new physics. Even if the pure QED contribution is by far the largest one, the experimental precision has reached a level where electroweak and hadronic contributions become important. The discrepancy has withstood the intense scrutiny so far which might indicate that other mechanisms could be contributing as well. Relaxing $T(CP)$-reversal symmetries can lead to additional contributions into both magnetic and electric dipole moments.

The primary effect of allowing for $P/T$-symmetry violation is the appearance of electric dipole moment (EDM). Additionally, a composite system (atoms and molecules) could have $P/T$-odd polarizabilities which are originated by a variety of symmetry-violating interactions. Conventionally, it is assumed that ``simpler'' particles (electron, muon, $\dots$) might have only EDMs as extensions of standard approach. It was suggested in \cite{baryshevsky_time-reversal-violating_2004}  that all types of particles (atom, molecule, nucleus, neutron, electron) can also have an additional property similar to $P/T$-odd polarizabilities (in addition to nonzero EDMs). However, a specific implementation of this idea for fermions was missing; this work fills this gap. 

If a system has $T$-odd polarizabilities and susceptibilities \cite{baryshevsky_time-reversal-violating_2004, baryshevsky_high-energy_2012}, see also \cite{ravaine_atomic_2005, kozlov_proposal_2006},  it can generate a  magnetic field upon applying the electric one; vice versa, an electric field can be generated by the magnetic one \cite{baryshevsky_time-reversal-violating_2004, baryshevsky_high-energy_2012}. As a direct  consequence of admitting nonzero $P/T$-odd polarizabilities, the magnetic and electric moments that are evaluated without these polarizabilities do acquire additional contributions. Saying in other words, interpreting the measured values of magnetic and electric moments might require taking account of such polarizabilities.

To explain how $T$-odd polarizabilities mix electric and magnetic moments, consider a particle (atom, nucleus, nucleon, electron, $\dots$) with a permanent magnetic moment, for example. It normally means that there exists an electric current that creates this moment as well as a magnetic field inside and around such a  particle. Without symmetry-violating polarizabilities, the contributions from these fields into the system energy (which are defined by corresponding moments) are cleanly separated from each other. Now, if the particle possesses a $T$-odd susceptibility (polarizability), this magnetic field will induce some electric field which magnitude is proportional to this polarizability; thus it effectively induces some electric dipole moment \cite{baryshevsky_search_2020}. Specifically, it can be shown as \cite{baryshevsky_time-reversal-violating_2004, baryshevsky_high-energy_2012}
\begin{equation}\label{eq604}
	d = d_0 + \mu \,\chi_{\text{Todd}}\,,
\end{equation}
where $d_0$ is the electric dipole moment evaluated without taking account of $T$-odd polarizability, $\mu$ is the particle magnetic moment, and $\chi_{\text{Todd}}$ is the $T$-odd susceptibility. The latter quantity is proportional to the $T$-odd polarizability normalized by some characteristic volume (occupied by the particle charge distribution).  Again, the particle magnetic moment means that there is a magnetic field (inside the particle) of order of $\mu/R^3$ where $R$ is a characteristic particle size. This field indices an additional electric dipole moment $\mu \chi_{\text{Todd}}\sim \mu \beta_{odd}/R^3 \sim d_{ind}$ by means of $T$-odd polarizability $\beta_{odd}$, see \cite{baryshevsky_time-reversal-violating_2004}.

By reciprocity, the EDM \eqref{eq604} creates an electric field which, in turn, creates a magnetic field and additional contribution into the magnetic moment which magnitude is again proportional to the $T$-odd polarizability. Hence, taking into account first the induced electric field, then the magnetic field that is induced by it, the additional contribution into the magnetic moment must be proportional to the square of $T$-odd polarizability  or the quantity that plays its role, see \eqref{eq128} below or \cite{baryshevsky_search_2020}.

 The squared Dirac equation allows to introduce the pseudoscalar quantity,  which plays the role of $T$-odd polarizability, into the spin motion equation \cite{baryshevsky_pseudoscalar_2020,porshnev_electric_2021}. Since such a polarizability mixes magnetic and electric moments, both problems, the search for EDM  and measurements of anomalous magnetic moment (AMM), are getting directly connected.  

In several recent studies \cite{baryshevsky_pseudoscalar_2020, baryshevsky_search_2020,porshnev_electric_2021}, we developed the phenomenological model which introduces the pseudoscalar corrections into the spin motion of fermions. The  approach is based on the squared Dirac equation with both AMM and EDM included.  The underlying physical hypothesis is the assumption that free fermions possess a small nonzero pseudoscalar $q_0=i\bar{\psi}\gamma^5\psi$.  This real-valued quantity is both $P$- and $T$-odd \cite{berestetskii_quantum_1982,sachs_physics_1987}. Bare charges are reduced by vacuum polarization which is one of the radiative corrections that is predicted by the field theory.  Effectively, a charged fermion is viewed as the core charge surrounded by the coat of virtual pairs, including its antifermions. The idea that the polarization of vacuum screening cloud around a bare charge should be taken into account in phenomenological models is not new \cite{baryshevsky_phenomenon_1999, baryshevsky_rotation_2008}, see also selected works on nucleon polarizability \cite{baryshevsky_high-energy_2012, baym_elementary_2016}. A magnetic field inside particle (estimated for a dipole with muon Compton length) could be enormous, be an order of $10^{16} G$. A virtual electron or muon, from the muon polarization coat, would have the cyclotron frequency  up to several dozens of $MeV$. Therefore, effects related to the vacuum polarization coat could be significant.  However,  a specific implementation of how to include such effects into phenomenological models was missing. As we discussed in \cite{baryshevsky_pseudoscalar_2020}, the inclusion of static pseudoscalar density into the phenomenological model is our attempt to take account of the vacuum polarization cloud within the single-particle representation. 

The squared Dirac equation has been viewed as an alternative to the regular first-order one  \cite{bethe_quantum_1957, berestetskii_quantum_1982, baier_radiative_1972, rafanelli_classical_1964, strange_relativistic_1998,bagrov_squaring_2018}. It properly recovers the term with $g$-factor and is frequently used in various applications \cite{berestetskii_quantum_1982}. Since it also recovers the Bargmann-Michel-Telegdi (BMT) equation in the quasiclassical limit \cite{rubinow_asymptotic_1963,rafanelli_classical_1964} and the Pauli term for the nonrelativistic motion \cite[p. 120]{berestetskii_quantum_1982}, it can be seen as equally confirmed by atomic phenomenology as the original Dirac equation.  To ensure the mathematical equivalency between two types of Dirac equations, only solutions of second-order equation that also satisfy the first-order equation  are conventionally chosen \cite[p. 120]{berestetskii_quantum_1982}. However, in doing so, we will end up with exactly the same situation as with the regular Dirac equation. The solutions of squared equation that are filtered in such a way will have zero pseudoscalar for the field-free case; we do not gain anything new then.  Therefore, we will not impose this additional requirement on the solutions of squared Dirac equation; it admits new types of solutions compared to the regular approach.
\\

\noindent\textbf{Main hypothesis and new physics:} Next, we consider the squared Dirac equation which allows to introduce the pseudoscalar density into the spin equation. In turn, it allows to derive the contribution of $T$-odd polarizability into both EDM and AMM that we discussed above directly within the formalism of squared Dirac equation. Following our previous work \cite{baryshevsky_pseudoscalar_2020}, it is convenient to define the new parameter
\begin{equation}\label{eq251}
	\beta = \frac{q_0}{r_0}= \eval{\frac{i\bar{\psi}\gamma^5\psi}{\bar{\psi}\psi}\,}_{\text{free}}
\end{equation}
where both quantities $q_0$ and $r_0$ are pseudodensity and density of free fermions. This parameter is both $P$- and $T$-odd and can play a role similar to the $P/T$-odd polarizability. 

Upon reviewing the available experimental data for both AMM and EDM problems within the same phenomenological approach, the following prediction can be made. Higher precision measurements of EDMs will continue yielding null results, while similar higher precision measurements of AMMs will continue confirming the gap against corresponding theoretical evaluations. The physical reason is the conversion of nonspherical electric moment into the additional magnetic anomaly by means of pseudoscalar density which plays a role $P$- and $T$- odd polarizability. This prediction is the extreme case among other possible scenarios that were reviewed in  \cite{baryshevsky_pseudoscalar_2020}. 

The $(g-2)$ discrepancy has withstood the intense scrutiny for quite some time already; however it is not a foregone conclusion that it could not be resolved without new physics one day \cite{blum_theory_2021}. The evaluation of hadronic vacuum polarization is still evolving; the estimates differ substantially between different teams \cite{borsanyi_et_al_hadronic_2018,giusti_light-quark_2018, davies_et_al_hadronic-vacuum-polarization_2020, davier_new_2020}.  Anyway, following the latest review \cite{aoyama_anomalous_2020} and new experimental data \cite{abi_et_al_measurement_2021}, the discrepancy currently stands at $(3.7-4.2)\sigma$. We have shown here that there exists the $T$-odd contribution into AMM (that is proportional to $\beta^2$) in addition to the conventional contributions that have been considered in the literature so far. Hence the magnetic anomaly is given as $a_{Teven} + a_{Todd}$ where the second term depends on the pseudoscalar $\beta$ from \eqref{eq251}. This additional contribution can possibly explain the muon $g-2$ discrepancy.
\\

The paper is organized as follows. In the next section, we briefly outline  the derivation of conventional  spin motion equation in weak external fields. Next, we recap the formalism based on the squared Dirac equation with both AMM and EDM terms included. Lastly, we discuss  contributions of $T(CP)$-odd pseudoscalar into both magnetic and electric dipole moments. The analysis of latest $(g-2)$ data for muons and electrons does not contradict the potentially significant contributions of pseudoscalar corrections into the magnetic anomaly.

\section{Conventional approach}
Spin motion  equations can be derived or justified in many different ways.  A large literature has been dedicated to this topic since the BMT equation was published 60+years ago \cite{bargmann_precession_1959}; the topic itself goes back to the Frenkel's work \cite{frenkel_elektrodynamik_1926}.  For homogeneous weak fields and the quasiclassical motion \cite{mane_spin-polarized_2005}, a consistent derivation leads to the BMT equation which can be upgraded with both anomalous magnetic and electric dipole moments. 

To improve the accuracy of regular Dirac equation, it is commonly upgraded \cite{foldy_electromagnetic_1952,sokolov_relativistic_1974, bagrov_dirac_2014}  by adding two field-dependent terms 
\begin{equation}\label{eq2}
	\qty[i \slashed \partial -  e \slashed A-m - (\frac{a e}{2m}+id \gamma^5)\frac{\sigma^{\mu\nu}}{2}F_{\mu\nu}] \psi = 0\,, 
\end{equation}
where $a$ and $d$ are the anomalous magnetic and electric dipole moments.  It is convenient to eliminate $\gamma^5$ from \eqref{eq2} by introducing the single tensor coefficient
\begin{equation}\label{eq40}
	\qty(i \slashed \partial -  e \slashed A-m - \frac{1}{2}b_{\mu\nu}\sigma^{\mu\nu}) \psi = 0\,, 
\end{equation}
which is defined by
\begin{equation}\label{eq69}
	b_{\mu\nu} = \frac{a e}{2m}F_{\mu\nu}-d\tilde F_{\mu\nu}\,,
\end{equation}
here the dual field  tensor $\tilde F^{\mu\nu}$ is defined as  $\varepsilon^{\mu\nu\rho\sigma}F_{\rho\sigma}/2$. Hence, the minimal coupling is extended by adding two terms stemming from the additional fermion form-factors.

The step-by-step WKB-based derivation of spin equation with AMM from equation \eqref{eq2} is given in \cite{rafanelli_classical_1964}, and with both AMM and EDM terms in \cite{baryshevsky_search_2020, porshnev_electric_2021}. It is given by
\begin{equation}\label{eedm_21}
	\dv{s^\mu}{\tau} 
	=\frac{ge}{2m} F^{\mu\nu}s_\nu+\frac{a e}{m} s^\rho F_{\rho\nu}u^\nu  u^\mu
	- 2d\qty( \tilde F^{\mu\nu}  s_\nu+  s^\rho \tilde F_{\rho\nu}  u^\nu u^\mu ) 
\end{equation}
where the spin and velocity vectors are defined as $s^\mu= (\bar{\psi}\gamma^5\gamma^\mu\psi) / (\bar{\psi}\psi)$ and $u^\mu= (\bar{\psi}\gamma^\mu\psi) / (\bar{\psi}\psi)$ respectively, $\tau$ is proper time,  and  $2a = g-2$. 

There exists the well known connection between the BMT-like equation \eqref{eedm_21} and the Thomas equation \eqref{eq401} which describes the spin precession in the laboratory frame \cite{nelson_search_1959,  berestetskii_quantum_1982, derbenev_polarization_1973, fukuyama_searching_2012}. These equations are self-consistent within this approximation order, if the fermion motion is driven by the regular Lorentz-Maxwell force. The evolution of spin three-vector $s_i$ is given by
\begin{equation}\label{eq401}
	\dv{\textbf{s}}{t} = \textbf{s}\cross\Omega\,,
 \end{equation}
where  
the precession frequency is given by
\begin{multline}\label{eq83}
	\Omega = \frac{e}{m}\qty[\qty(a+\frac{1}{\gamma}) \textbf{B}-\frac{a\gamma}{\gamma+1}  (\textbf{v}\cdot\textbf{B}) \,\textbf{v}-\qty(a+\frac{1}{\gamma+1})\textbf{v}\cross\textbf{E} ]\\[1ex]
	 + 2d \qty[ \textbf{E} - \frac{\gamma}{\gamma+1} (\textbf{v}\cdot\textbf{E}) \,\textbf{v}+ \textbf{v}\cross \textbf{B} ]\,.
\end{multline}
The spin three-vector $\textbf{s}$ is given in the rest frame, while the fields are in the laboratory frame, and $\gamma$ is the Lorentz factor.

In a non-relativistic frame the precession frequency becomes 
\begin{equation}\label{eq90}
	\Omega_{nr} = \frac{e}{m}\Big[\big(a+1\big)\textbf{B}-\big(a+\frac{1}{2}\big)\textbf{v}\cross \textbf{E}\Big] + 2d \big( \textbf{E}+\textbf{v}\cross\textbf{B} \big)\,.
\end{equation}  
In the semiclassical approximation, we assume that the space part of wave packet moves along an orbit that is  a solution of corresponding Hamilton-Jacobi equation. More accurately, such orbits can be described by eigenfunctions of scalar part of Hamiltonian, since particle trajectories are only weakly influenced by spin in this approximation. The spin evolution along such trajectories is given by the spin precession \eqref{eq401} with frequency \eqref{eq83} or \eqref{eq90}. {color{blue} The applicability conditions of this approximation are discussed in \cite{berestetskii_quantum_1982, mane_spin-polarized_2005}.}

\section{Squared Dirac equation}
Consider now the equation \eqref{eq2} where we re-label coefficients $d$ and $a$ as $d_0$ and $a_0$ respectively; they are quantities that are used in conventional theories which do not take account of $T$-odd polarizability. Next, following the traditional approach \cite[p.  119]{berestetskii_quantum_1982}, the squared Dirac equation is obtained by applying the adjoint operator to \eqref{eq40}
\begin{equation}\label{eq105}
	(i \slashed \partial -  e \slashed A_\mu+m + \frac{1}{2}b^0_{\mu\nu}\sigma^{\mu\nu})(i \slashed \partial -  e \slashed A-m - \frac{1}{2}b^0_{\mu\nu}\sigma^{\mu\nu}) \psi=0 \,,
\end{equation}
where $b^0_{\mu\nu}$ is defined with \eqref{eq69} by means of $d_0$ and $a_0$. The step-by-step WKB-based derivation of spin equation which follows from \eqref{eq105} is given in \cite{baryshevsky_search_2020, porshnev_electric_2021}. The key difference from the previous section is allowing a nonzero pseudoscalar $q\ne 0$ or, equivalently, admitting $\beta\ne 0$. Remarkably, we obtain exactly the same BMT-like equation which we repeat here as 
\begin{equation}\label{eq132}
	\dv{s^\mu}{\tau} 
	=\frac{g e}{2m} F^{\mu\nu}s_\nu+\frac{a e}{m} s^\rho F_{\rho\nu}u^\nu  u^\mu
	- 2d \qty( \tilde F^{\mu\nu}  s_\nu+  s^\rho \tilde F_{\rho\nu}  u^\nu u^\mu ) 
\end{equation}
where again $2a = g-2$. However, the moments  (or coefficients in front of field-dependent terms in the above equation) are now given as  
\begin{equation}\label{eq603}
	a = a_0 + d_0 \,\frac{2m}{e}\,\beta\,,	\qquad\qquad d = d_0 -  a_0\,\frac{ e}{2m}\,\beta\,.
\end{equation}
As we discussed in the introduction, particles are typically assumed to have no $T$-odd polarizability in addition to EDM. Assuming instead that such additional polarizabilities are nonzero leads to corrections to magnetic and electric moments that are evaluated without $T$-odd polarizabilities. Physical reasons behind the appearance of $T$-odd polarizabilities in a phenomenological models are $T$-odd interactions of fermions with some other background field that might be of pseudoscalar nature. A specific mechanism behind the appearance of $T$-odd polarizability of quasiclassical fermions is outside the scope of this work; we can however point to the analogy with atomic systems where certain types of hypothetical interactions between electrons and nucleus nucleons lead to appearance of  such polarizabilities \cite{baryshevsky_high-energy_2012}.  The form of the additional contributions to magnetic and electric moments (evaluated without $T$-odd polarizabilities) could be guessed from the dimensional analysis; however, it would not be possible to derive numerical factors in front of these contributions. The formalism developed here allows to derive the exact form of additional $T$-odd contributions and additionally discover that the Dirac pseudoscalar can be viewed as an effective $T$-odd polarizability of fermion.

Inspecting the coefficients \eqref{eq603}, we see that the pseudoscalar mixes the magnetic and electric moments in them. However, since $\abs{d_0}\ll \abs{a_0e/m}$ and $\abs{\beta}\ll 1$, the correction to $a$ is of second order of smallness.   It is similar to $T$-odd  and $P$-odd atomic polarizabilities that are originated by symmetry-violating interactions of atomic electrons with nucleus nucleons   \cite{baryshevsky_high-energy_2012}. The scalar parts of these polarizabilities are pseudoscalar quantities which mix electric and magnetic contributions into the system energy. It is quite a typical situation in systems where symmetry-violating interactions are allowed, as we discussed in \cite{baryshevsky_search_2020}. In the scenario where the EDM is screened $(d_0\sim \beta)$, see below, the pseudoscalar correction to the magnetic moment is of order of $\beta^2$ which physical origin was discussed in the introduction. 

The remaining task for this section is to re-write the spin equation for the laboratory frame. Since \eqref{eq132} has the same functional form as the original BMT equation \eqref{eedm_21}, the known connections and relationships \cite{berestetskii_quantum_1982, fukuyama_derivation_2013}  between two equations  can be directly used. The spin precession equation has the same form \eqref{eq401} 
\begin{equation}\label{eq402}
	\dv{\textbf{s}}{t} = \textbf{s}\cross\Omega^\beta\,,
\end{equation}
where three-vectors of spin $\textbf{s}$ and velocity $\textbf{v}$ keep their conventional definitions.  The precession frequency is again given by
\begin{multline}\label{eq86}
	\Omega^\beta =\frac{e}{m} \Big[\big(a+\frac{1}{\gamma}\big) \textbf{B}-\frac{a\gamma}{\gamma+1}  (\textbf{v}\cdot\textbf{B}) \,\textbf{v}-\big(a+\frac{1}{\gamma+1}\big)\textbf{v}\cross\textbf{E}) \Big]\\[1ex]
			 + 2d \qty[ \textbf{E} - \frac{\gamma}{\gamma+1} (\textbf{v}\cdot\textbf{E}) \,\textbf{v}+ \textbf{v}\cross \textbf{B} ]\,.
\end{multline}
where the  coefficients are now defined as
in \eqref{eq603}. We see that all terms keep their original physical meaning and form (including Thomas precession), while both magnetic and electric moments are adjusted with pseudoscalar corrections.

The nonrelativistic precession frequency follows from \eqref{eq86} as
\begin{equation}\label{eq87}
 	\Omega^\beta_{nr} = \frac{e}{m} \Big[\big(a+1\big) \textbf{B}-\big(a+\frac{1}{2}\big)\textbf{v}\cross\textbf{E}) \Big]\\[1ex]
			 + 2d \big( \textbf{E} + \textbf{v}\cross \textbf{B} \big)
\end{equation}
where we dropped terms with powers of velocity equal or higher than two. Here the moments are given by 
\begin{equation}\label{eq133}
	a= a_0 + d_0 \,\frac{2m}{e}\,\beta\,,	\qquad\qquad d = d_0 -  a_0\,\frac{ e}{2m}\,\beta\,,
\end{equation}
which are the same expressions as \eqref{eq603};  they are given here again to emphasize that the same expressions are valid for the nonrelativistic case in this approximation order.

As a final comment for this section, we would like to mention that we obtained the additional contribution into the EDM, see the second equation in \eqref{eq133},  in the form $a_0\,\mu_B\,\beta$, where $\mu_B= e/2m$ is the Bohr magneton. A similar form was discussed before \cite{baryshevsky_time-reversal-violating_2004}, see also the discussion around equation \eqref{eq604} regarding the influence of magnetic field on induced EDM. Later, this form was also used in \cite{commins_electric_2009, engel_electric_2013}, however in the very different context. The electron EDM was parameterized there as 
\begin{equation}\label{eedm_5}
	d_e  =  a_0\, \mu_B\, \lambda \,,
\end{equation}
where the dimensionless coefficient $\lambda$ combines model-specific mass ratios, mixing angles, and couplings. This \emph{ad~hoc} form was proposed based on purely dimensional grounds in \cite{engel_electric_2013}
	 \footnote{In parameterized forms, the anomalous moment $a_0$ is often replaced by the leading (Schwinger) term $\alpha/2\pi$.}.
The form \eqref{eedm_5} was known since \cite{baryshevsky_time-reversal-violating_2004}, it is derived here strictly from the squared Dirac equation; more, the dual contribution into the AMM, see the first equation in \eqref{eq133}, has not been known so far.  Comparing the \emph{ad~hoc} form \eqref{eedm_5} with our equations \eqref{eq133}, we can match the parameters as $\lambda a_0\sim \beta a_0 \sim \chi$ which is the $T$-odd susceptibility from \eqref{eq604}.  

\section{Evaluating contribution of T-odd pseudoscalar into AMM and EDM} 
Let us label experimentally measured moments as $a^{\text{exp}}$ and $d^{\text{exp}}$ respectively. The treatment of experimental data is based on many assumptions. While the new $T(CP)$- symmetry violating effects are extensively studied in the field theory \cite{chupp_electric_2019}, the contributions into EDM and AMM that are derived there do not take account of $T$-odd polarizability.

Several scenarios are possible depending on the magnitudes of dipole moments and ratio of densities $\beta$, see the table.  
		\begin{table}[H]
									\centering %
						\begin{tabular}{@{}cccclll@{}}
					& $d^{\text{exp}}$ 	& \scalebox{0.9}{$\Delta a=a^{\text{exp}}-a_0$} & $\beta$ & Comment\\[2pt]\toprule 
			$1$ 	& $0$			& $\phantom{\ne}\,0$ & $\phantom{\ne}\,0$& No New Physics (NP) \\[2pt]
		$2$ 	& $d^{\text{exp}}$	& $\phantom{\ne}\,0$ & $\phantom{\ne}\,0$& NP, conventional model \\[2pt]
		$3$ 	& $d^{\text{exp}}$	 &$\ne0$ 			& $\ne0$ & NP, mixed case, new model\\[2pt]
		$4$ 	& $0$	& $\abs{a^{\text{exp}}}>\abs{a_0}$ 			& $\ne 0$ & NP, screened EDM, new model
		\end{tabular} 
\end{table}
\noindent In this context, the coefficient $a_0$ is associated with the theoretical value $a^{\text{th}}$, since the existing theoretical evaluations of both magnetic and electric moments do not take into account the possibility of nonzero $\beta$. The conventional models are still adequate in describing experiments at the current level of accuracy (Case 1 and 2).  If the $g-2$ discrepancy persists, then it signals that $\beta\ne 0$. It means  the pseudoscalar corrections must be taken into account according to 
\eqref{eq133}. The corrections offset both magnetic and electric moments that are evaluated without taking into account the $\beta$-correction. Case 3 favors experiments with the heaviest fermions, since the relative strength of corrections scales as $m^{-2}$, see \cite{baryshevsky_search_2020}.

The most restrictive scenario (Case 4), which is also the most predictive, is the assumption that the EDM is screened by the vacuum polarization cloud.  It follows then from \eqref{eq133} that
\begin{equation}\label{eq131}
	d^{\text{exp}}\approx0\qquad\to\qquad d_0  \approx \frac{ae}{2m} \beta\,.
\end{equation}
Substituting it into the first equation in \eqref{eq133}, for $\gamma\to 1$ for example,  we see that the  model yields the additional contribution $\beta^2 a_0$  into the magnetic moment 
\begin{equation}\label{eq128}
	a^{\text{exp}} \approx  a_0 \big(1 + \beta^2\big)\,.
\end{equation}
Hence, the new model predicts that the experimental data will exceed the existing theoretical values
\begin{equation}\label{eq97}
	\abs{ a^{\text{exp}}}   > \abs{ a_0}\,.
\end{equation}
The inequality holds independently of signs of magnetic anomaly $a_0$, the static pseudoscalar $q_0$ (or $\beta$ since $r_0>0$), and dipole moment $d_0$. Clearly, the underlying assumption is that both experiment and theoretical evaluation achieved a level where the pseudoscalar correction becomes the primary source of difference $(a^{\text{exp}}- a_0)$.

Physically, the prediction \eqref{eq131}-\eqref{eq97}, can be explained as follows. The vacuum polarization cloud around the bare charge screens the particle EDM, however the electric screening leads to the  increased magnetic anomaly. Saying otherwise, the unbalanced electric charge distribution gets converted into the additional magnetic moment by means of pseudoscalar density which plays a role similar to $P$- and $T$-odd polarizabilities, see \cite{baryshevsky_search_2020}.  

These considerations allowed to suggest the new explanation for the $(g-2)$ muon discrepancy between the best experimental $(a_\mu)^{\text{exp}}$ and latest theoretical  $a_\mu$ values \cite{aoyama_anomalous_2020}
\begin{equation}\label{eq99}
	(a_\mu)^{\text{exp}} - a_\mu = 2.8 \times 10^{-9} \,.
\end{equation}
The efforts to address remaining theoretical uncertainties in \eqref{eq99} have been relentless; the most significant uncertainties are contributions from higher-order hadronic light-by-light diagrams \cite{hoferichter_dispersion_2018,colangelo_longitudinal_2020, blum_et_al_hadronic_2020} and hadronic vacuum polarization \cite{borsanyi_et_al_hadronic_2018,giusti_light-quark_2018, davies_et_al_hadronic-vacuum-polarization_2020, davier_reevaluation_2017, davier_new_2020},  for which the teams seem to be making good progress recently.  Still, the discrepancy remains and currently stands at  $4.2\sigma$ level \cite{abi_et_al_measurement_2021}. Assuming now that the main factor behind the discrepancy \eqref{eq99} is the $\beta$-correction \eqref{eq128} allows to determine both $d_\mu$ and $\beta_\mu$, see their estimates in \cite{baryshevsky_pseudoscalar_2020}.

However, the electron AMM data that were available to us by the time of preparing \cite{baryshevsky_pseudoscalar_2020} showed that $(a_e)^{\text{exp}} - a_e< 0$; clearly, it contradicts to the model prediction \eqref{eq97}. We commented in \cite{baryshevsky_pseudoscalar_2020} that the only way to resolve this conflict within the developed model and under the assumption that electron EDM is screened is to accept that the currently available electron data had not reached the accuracy level where the $\beta$-correction becomes dominant. Our model does not include any other parameters that can be fitted to reverse the inequality \eqref{eq97}; it is quite restrictive in this case. Since that time,  the new value of fine structure constant has become available \cite{morel_determination_2020}. It has allowed to reevaluate the discrepancy between experimental and theoretical values of electron magnetic anomaly \cite{andreev_et_al_constraints_2021} as
\begin{equation}\label{eq129}
	(a_e)^{\text{exp}} - a_e = 4.8 \times 10^{-13}\,.
\end{equation}
Now, the difference in electron AMM values between the experiment and theory agrees with  our model prediction.

The standard model predicts the very small EDM values of around $10^{-39}e\cdot cm$ which was recently re-estimated by taking into account the one-loop vector meson contributions \cite{yamaguchi_large_2020, yamaguchi_quark_2021}. Assuming that the latest estimate \eqref{eq129} withstands the scrutiny, our phenomenological model is able to explain this difference and extract both $d_e$ and $\beta_e$. A much higher value of intrinsic electron EDM  is expected if our approach is valid. It might be difficult to measure directly, since we expect that it is screened.

A few final comments. The new results that we outlined in this short paper are summarized as follows
\begin{itemize}
	\item Solutions of squared Dirac equation include the $\beta$-parameter which can play a role of $T$-odd polarizability
	\item Analysis of experimental data does not contradict the potentially significant contributions of pseudoscalar corrections into the magnetic anomaly of fermions. 
\end{itemize}
If both AMM and EDM problems are considered in the single framework, the following picture is emerging. Increasing simultaneously the accuracy of experimental and theoretical results, the EDM measurements across the board continue yielding null results, while the gaps between measured and theoretical values of AMM emerge. These gaps are predicted to have the clear signature, $\abs{a^{\text{exp}}} > \abs{a^{\text{th}}}$, and  are now observed for both electrons and muons. 
	\footnote{\color{gray} In this statement, we assume that experimental data are extracted by using conventional phenomenology that does not include pseudoscalar corrections, hence the gaps.}
In this connection, an experiment to measure the magnetic anomaly of $\tau$-leptons is predicted to the find the similar gap and hence has an additional interest since it could confirm the universality of \eqref{eq97}. It is explained by the presence of vacuum polarization around the bare charges that screens the intrinsic EDM which in turn leads to the  increased magnetic anomaly. Saying differently, the nonspherical electric charge distribution gets effectively converted into the additional magnetic moment. 

\section*{Acknowledgments}
We thank B.~Malaescu for pointing the reference \cite{morel_determination_2020} to us, and R.~Talman for valuable discussion of our results which helped to clarify the presented material.



\end{document}